\documentclass[useAMS,usenatbib]{mn2e}

\usepackage{graphicx}
\usepackage{amssymb}

\newcommand{\apj}{ApJ}

\newcommand{\aap}{A\&A}

\newcommand{\mnras}{MNRAS}

\newcommand{\kms}{km~s$^{-1}$}

\newcommand{\HI}{H{\sc i}}
\newcommand{\HII}{H{\sc ii}}
\newcommand{\sunn}{$_{\odot}$}
\DeclareRobustCommand{\ion}[2]{%
\relax\ifmmode
\ifx\testbx\f
{\mathrm{#1\,\textsc{#2}}}\else
{\mathrm{#1\,\mathsc{#2}}}\fi
\else\textup{#1\,{\mdseries\textsc{#2}}}%
\fi}

\title[Eridanus void LSBDs J0015+0104 and J2354--0005]
{Properties of the most metal-poor gas-rich LSB dwarf galaxies
SDSS J0015+0104 and J2354--0005 residing in the Eridanus void}
\author[S.A. Pustilnik, J.-M. Martin, Y.A. Lyamina, A.Y. Kniazev
]
{S.A. Pustilnik,$^{1,6}$\thanks{E-mail: sap@sao.ru (SAP)}
J.-M. Martin,$^{2}$
Y.A. Lyamina,$^{3}$
A.Y. Kniazev$^{4,5,7}$ \\
\rule{-4pt}{20pt}
$^1$ Special Astrophysical Observatory of RAS, Nizhnij Arkhyz,
  Karachai-Circassia 369167, Russia\\
$^2$ GEPI and Station de radioastronomie, Observatoire de Paris,
  5 place Jules Janssen, 92190 Meudon, France  \\
$^3$ Physics Department, Southern Federal University, Rostov-on-Don, Russia\\
$^4$ South African Astronomical Observatory, PO Box 9, 7935 Observatory,
   Cape Town, South Africa\\
$^5$ Southern African Large Telescope Foundation, PO Box 9, 7935 Observatory,
   Cape Town, South Africa \\
$^6$ Isaac Newton Institute of Chile, SAO branch, Nizhnij Arkhyz, Russia \\
$^{7}$ Sternberg Astronomical Institute, Lomonosov Moscow State University,
Moscow, Russia
}

\begin{document}

\label{firstpage}

\date{Accepted on 2013 April 2. Received on 2012 July 10}

\pagerange{\pageref{firstpage}--\pageref{lastpage}} \pubyear{2013}

\maketitle

\begin{abstract}

SDSS J0015+0104 is the lowest metallicity low surface brightness
dwarf (LSBD) galaxy known. The oxygen abundance in its \HII\ region SDSS
J001520.70+010436.9 (at $\sim$1.5~kpc from the galaxy centre) is
12+$\log$(O/H)=7.07 (Guseva et al.). This galaxy, at the distance
of 28.4~Mpc, appears to reside deeply in the volume devoid of luminous
massive galaxies, known as the Eridanus void. SDSS J235437.29$-$000501.6 is
another Eridanus void LSBD galaxy, with parameter 12+$\log$(O/H)=7.36
(also Guseva et al.).
We present the results of their \HI\ observations with the Nan\c {c}ay Radio
Telescope revealing their high ratios of $M$(\HI)/L$_{\rm B}$ $\sim$2.3.
Based on the Sloan Digital Sky Survey images, we derived for both
galaxies their radial surface brightness profiles and the main
photometric parameters.
Their colours and total magnitudes are used to estimate the galaxy stellar
mass and ages. The related gas mass-fractions, $f_{\rm g} \sim$0.98 and
$\sim$0.97, and the extremely low metallicities (much lower than for
their more typical counterparts with the same luminosity) indicate their
unevolved status. We compare these Eridanus void LSBDs with several
extreme LSBD galaxies residing in the nearby Lynx-Cancer void. Based on
the combination of all their unusual properties, the two discussed LSBD
galaxies are similar to the unusual LSBDs residing in the closer void.
This finding presents additional evidence for the existence in voids of a
sizable fraction of low-mass unevolved galaxies. Their dedicated search
might result in the substantial increase of the number of such objects in
the local Universe and in the advancement of understanding their nature.
\end{abstract}


\section[]{INTRODUCTION}
\label{sec:intro}

The modern cosmological CDM models of the large-scale structure and galaxy
formation, including the state-of-art N-body simulations, predict that
galaxy properties and evolution can significantly depend on their global
environment
\citep[e.g.,][and references therein]{Peebles01, Gottlober03, Hoeft06,
Hoeft10, Hahn07, Hahn09, Kreckel11b}. However, the role of the most rarefied
environment (typical of voids) in galaxy formation and evolution is not well
studied both theoretically and observationally.

The Lynx-Cancer void, one of the nearest, is situated in the sky region
well covered by the Sloan Digital Sky Survey (SDSS).
In \citet[][hereafter Paper~I]{PaperI} we described this void and the sample
of 79 galaxies residing inside it. Their absolute magnitudes
$M_{\rm B}$ range from --12.0 to --18.4 with the median value of --14.0.
In Paper~II \citep{void_OH} we presented the results of their gas
metallicity study.

In course of systematic study of dwarf galaxies in the Lynx-Cancer void,
we have already discovered half-dozen unusual objects, including DDO~68
\citep*{DDO68,IT07,DDO68_sdss}, J0926+3343 \citep{J0926} and other very
metal-poor low surface brightness dwarfs 
\citep[][LSBDs; hereafter Paper~III]{void_LSBD}. The recent 
Giant Metrewave Radio Telescope (GMRT)
mapping of a subsample of this void galaxies revealed two extremely gas-rich
blue dwarfs \citep{CP2013}.
In total, $\sim$10 per cent of the void dwarfs
show very unusual properties. They are very gas rich LSBDs with values of O/H
which are several times lower than expected for their $L_{\rm B}$ from
the relation "O/H versus $L_{\rm B}$"\, for a denser environment in the Local
Volume  (e.g. \citet{vZee06}).
Some of them have rather blue colours of outer parts that
correspond to ages of their main stellar population of $T \sim$1--4 Gyr.
The high concentration of unusual, "unevolved"\, dwarfs in this void
is a clear indication on the importance of void conditions for the
slow galaxy evolution and/or for the retarded galaxy and star formation (SF).

Recently  \citet{Guseva09} presented, among other new metal-poor galaxies,
the second most metal-poor galaxy SDSS J0015+0104, with 12+$\log$(O/H)=7.07.
This is a bona fide LSBD with a single faint \HII\ region on the
edge of the optical body.
For comparison, in the record-holder dIrr galaxy \mbox{SBS~0335--052W}
this parameter varies a little along the galaxy body around $\sim$7.0
\citep{Izotov09}.
We present here new \HI\ observations of J0015+0104 and the
analysis
of its SDSS $u,g,r,i$ images  along with the examination of its large-scale
environment. This LSBD appears to reside in a large region devoid
of luminous galaxies, known as the Eridanus void. We therefore 
compare its properties with those of several of the most metal poor LSBD
galaxies in the Lynx-Cancer void. We present also the similar study of
another very metal poor dwarf in the Eridanus void, SDSS
\mbox{J2354--0005}.
The paper is organised as follows. In Section~\ref{sec:obs} we briefly describe
observations and reduction of obtained data. Section~\ref{sec:results}
presents the results of observations and their analysis.
In Section~\ref{sec:dis} we discuss the results and their implications in a
broader context and summarise our conclusions.

\section[]{OBSERVATIONS AND DATA REDUCTION}
\label{sec:obs}

\subsection{\HI\ observations}

\HI\ observations with the
Nan\c {c}ay\footnote{The Nan\c {c}ay Radioastronomy Station is part of the
Observatoire de Paris and is operated by the Minist\`ere de l'Education
Nationale and Institut des Sciences de l'Univers of the Centre National
de la Recherche Scientifique.}
Radio Telescope (NRT) with a collecting area of
200$\times$34.5~$m^2$ are characterised by a half-power beam width of
3.7 arcmin~(east-west)$\times$22 arcmin~(north-south) at declination
$\delta$=0\degr\ (see also \verb|http://www.obs-nancay.fr/nrt|).
The data were acquired during 2011 June -- 2012 February, with the total time
on-source of $\sim$140~min. for J0015+0104 and $\sim$110~min -- for
J2354--0005. We used the antenna/receiver system F.O.R.T.
\citep[e.g.,][]{FORT}. The system
temperature was $\sim$35 K for the two circular polarisation outputs of the
receiver.   The gain of the telescope
was 1.5 K~Jy$^{-1}$ at declination $\delta$=0\degr.
The 8192-channel correlator was used covering the total bandwidth of
12.5 MHz. The total velocity range covered was about 2700~\kms, with
the channel spacing of 1.3~\kms\ before smoothing.
For more details on the NRT observations see the description in the
paper by \citet{NRT_07}.

The left and right polarisation spectra were calibrated and processed
independently and then averaged together. The error estimates were
calculated following to  \citet{Schneider86}. The baseline was  well fitted
by a third-order or lower polynomial  and was subtracted out.

The NRT full width at half maximum =22 arcmin corresponds 
to angular distances of $\pm$11 arcmin
in the NS direction. At distances of the target LSBDs of $\sim$30 Mpc, this
corresponds to linear separations of $\sim$100 kpc. No potential
confusung neighbours was found within the NRT beam.

\subsection{Imaging data from the SDSS}

\begin{figure*}
 \centering
 \includegraphics[angle=-0,width=15cm]{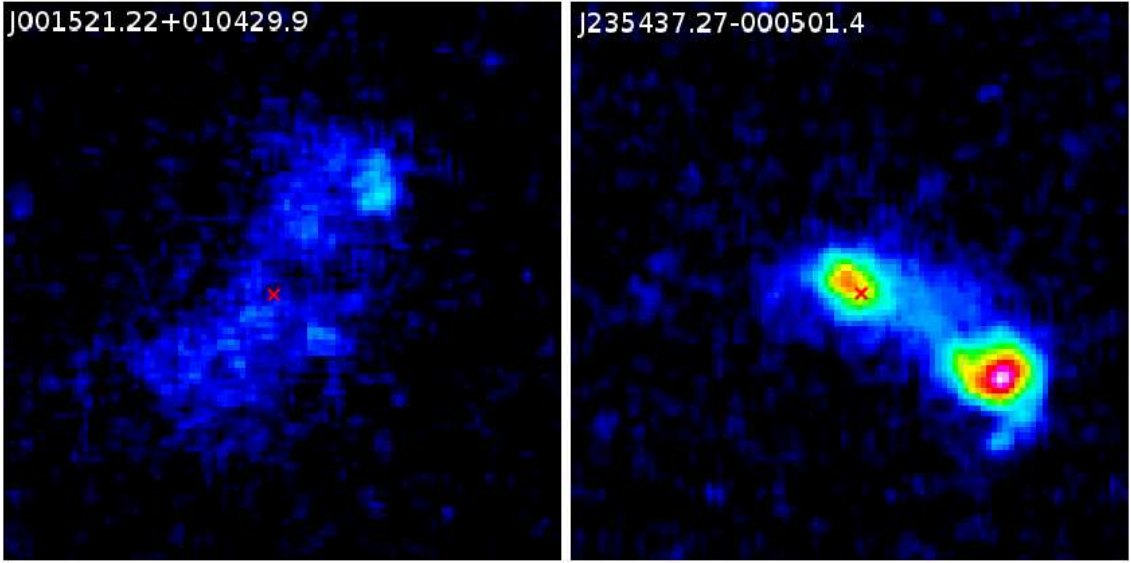}
  \caption{\label{fig:image}
Left-hand panel:
the $g$-filter image of SDSS J0015+0104 in conditional colours, centred
at the geometrical centre of the LSB body, marked by cross. The total size
of the field is $\sim$40 arcsec by $\sim$40 arcsec. North is up, East is
to the right.
Right-hand panel:
the similar image of SDSS J2354--0005. The neighbouring object at
  $\sim$10 arcsec to SW is a background galaxy with redshift of $\sim$0.165
   (Guseva et al. 2009). It was masked in making surface
   photometry of the studied galaxy.
}
\end{figure*}

The SDSS \citep{York2000} is well suited for
photometric studies of various galaxy samples due to its homogeneity, area
coverage and depth (SDSS Project Book\footnote{
http://www.astro.princeton.edu/PBOOK/}).
SDSS is an imaging and spectroscopic survey that covers about
one-quarter of the Celestial Sphere. The imaging data are collected in
drift scan mode in five bandpasses \citep[$u, \ g, \ r, \ i$, and $z$;]
[]{SDSS_phot} using mosaic CCD camera \citep{Gunn98}. An automated
image-processing system detects astronomical sources and measures their
photometric and astrometric
properties \citep{Lupton01,SDSS_phot1,Pier03} and identifies candidates for
multi-fibre spectroscopy. At the same time, the pipeline reduced SDSS data
can be used for making ones own photometry \citep[e.g.,][]{Kniazev04}
any project needs.
For our current study the images in the respective filters were retrieved
from the SDSS Data Release 7 \citep[DR7;][]{DR7}. The respective finding
charts with the field size of $\sim$40 arcsec are shown in
Fig.~\ref{fig:image}.

Since the SDSS provides users with the fully reduced images, the only
additional step we needed to perform (apart the photometry in round
apertures) was the background subtraction. For this, all bright stars
were removed from the images. After that, the studied object was masked and the
background level within this mask was approximated with the package
{\tiny AIP} from {\tiny MIDAS}.  In more detail the method and the related
programs are described in \cite{Kniazev04}.
To transform instrumental fluxes in apertures to stellar magnitudes, we
used the photometric system coefficients defined in SDSS for the used
fields. The accuracy of zero-point determination was  $\sim$0.01 mag in all
filters.

\section[]{RESULTS}
\label{sec:results}

\subsection[]{\HI\ parameters}

\begin{figure*}
  \centering
 \includegraphics[angle=-90,width=7.0cm, clip=]{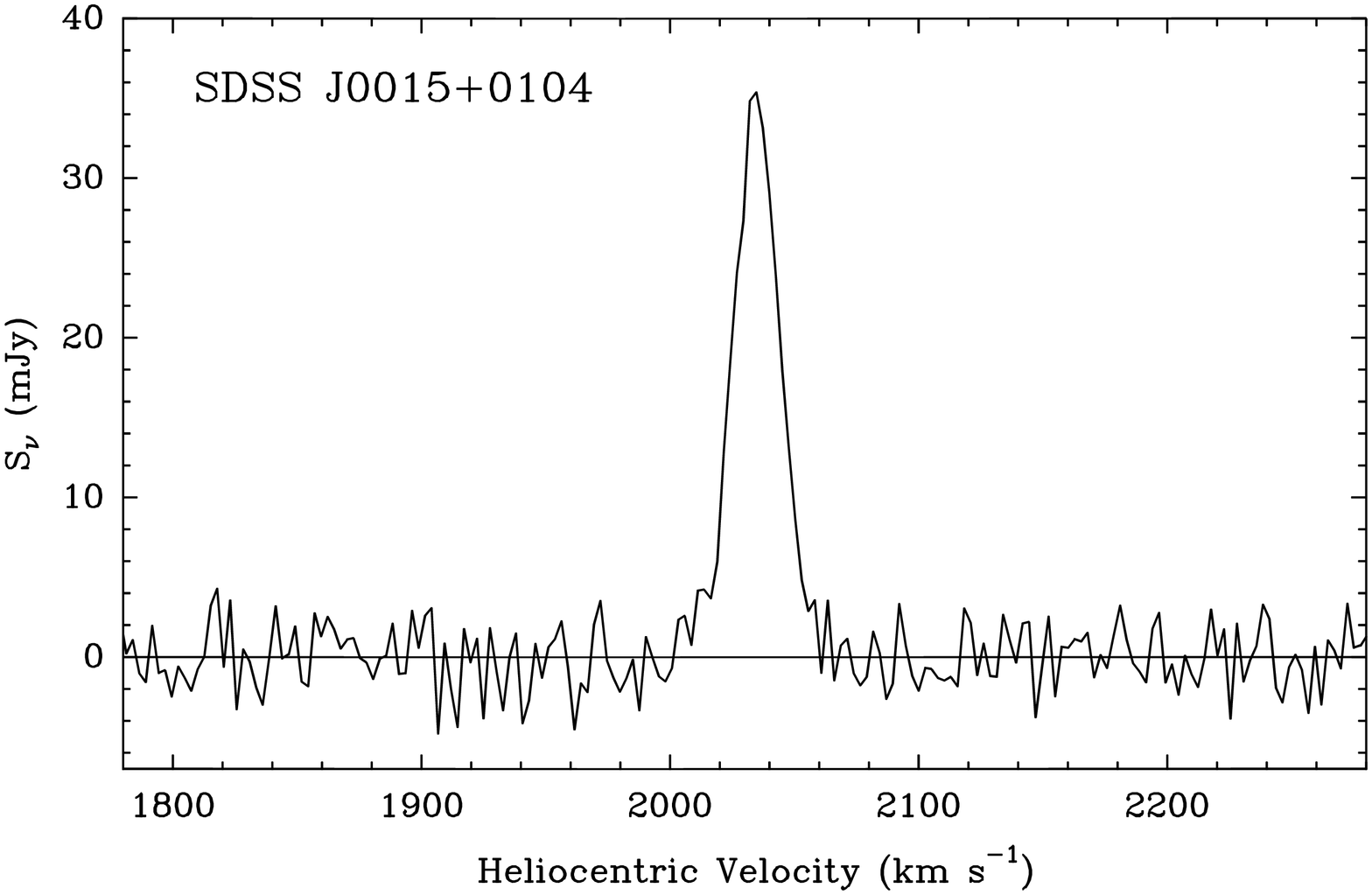}
 \includegraphics[angle=-90,width=7.0cm, clip=]{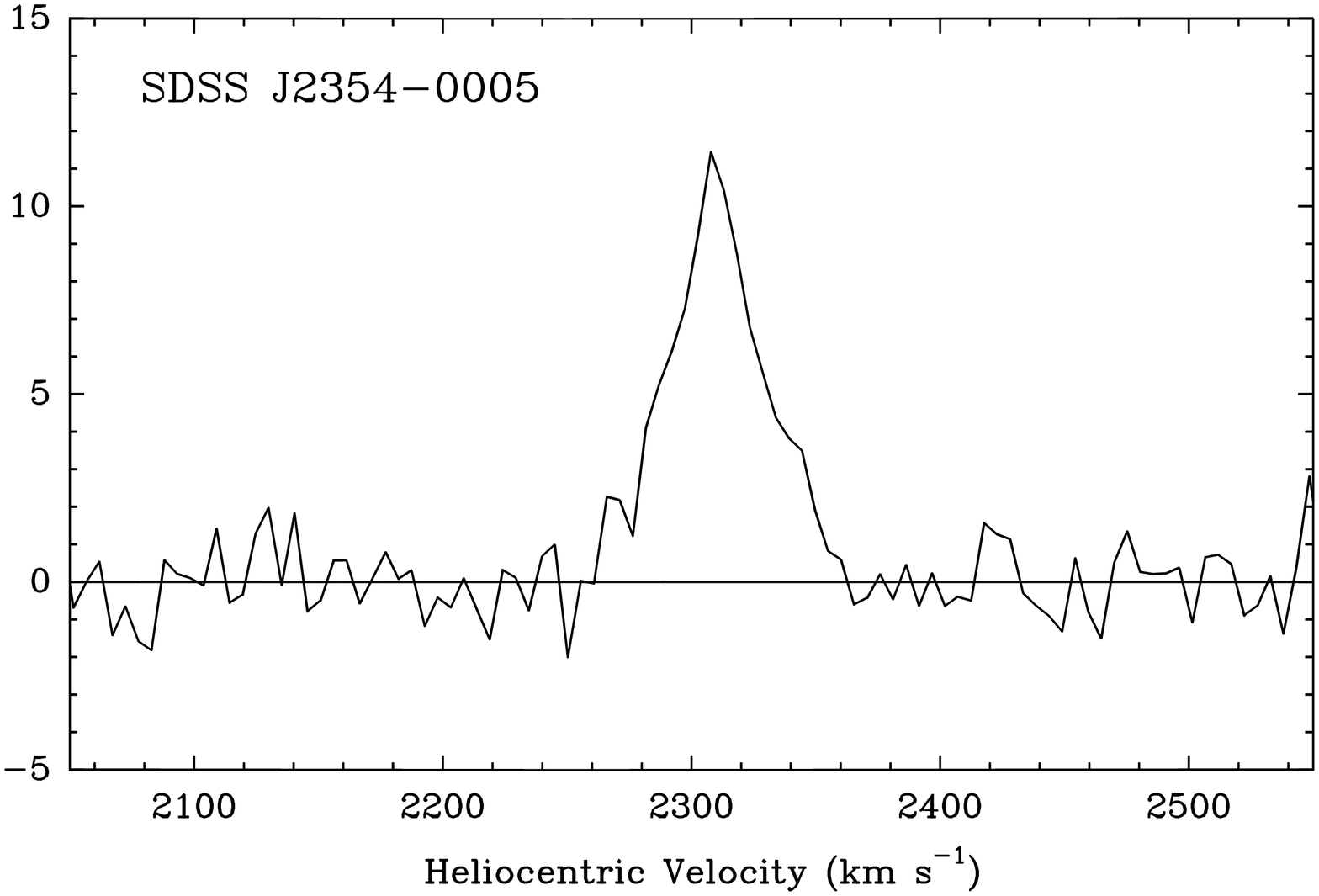}
  \caption{
The NRT profiles of \HI-line emission in galaxies SDSS J0015+0104 and
SDSS J2354--0005. The X-axis shows
radial heliocentric velocity in \kms. The Y-axis shows the galaxy flux density
in mJy.
}
	\label{fig:HI}
 \end{figure*}

The profiles of the 21 cm \HI-line emission at positions of SDSS J0015+0104
and J2354--0005 obtained with the NRT are shown in Fig.~\ref{fig:HI}. The high
S-to-N ratio narrow \HI\ profile of J0015+0104 looks as a simple  Gaussian
and its parameters are determined rather straightforwardly.
Its integrated \HI-line flux $F$(\HI)=0.81$\pm$0.04 Jy~\kms.  % S/N=27!
The central velocity of the  profile is 2035$\pm$3~\kms\
(in comparison to the SDSS emission-line value for the NW H{\sc ii}
region $V_{\rm opt}$=2066$\pm$64).
The profile widths are
W$_\mathrm{50}$=21.2$\pm$2~\kms\ and W$_\mathrm{20}$=29.4$\pm$3~\kms.

For SDSS J2354--0005, the profile is broader and its S-to-N ratio is
lower. Its integrated \HI-line flux $F$(\HI)=0.50$\pm$0.04~Jy~\kms.
The parameter $W_\mathrm{50}$=38.5$\pm$7.3~\kms.
The central velocity of \HI\ peak (2310$\pm$4~\kms)  is very close to
the SDSS optical velocity of $V_{\rm opt}$=2311~\kms\ and the independent
similar value of \citet{Guseva09}.

To estimate the galaxies' global parameters, we adopted the distance
$D$=28.4~Mpc for J0015+0104. This is based on the NED value with the Hubble
constant of 73~\kms~Mpc$^{-1}$, with the Virgo infall model, and
with correction down by 0.4 Mpc, accounting for the difference between the
NED V$_{\rm hel}$=2066$\pm$64~\kms\ and a new one of better accuracy,
derived from our \HI\ profile. Similarly, for J2354--0005 we adopted the
distance  $D$=32.1~Mpc. The respective scales are 138~pc and 156~pc in
1 arcsec.
The \HI\ mass of the galaxies is determined by the well-known relation for
optically thin \HI-line emission from \citet{Roberts69}.
This gives $M$(\HI)=(1.82$\pm$0.15)$\times$10$^{8}$~M\sunn\ for
J0015+0104 and (1.22$\pm$0.11)$\times$10$^{8}$~M\sunn\ for J2354--0005.

\subsection{Photometric properties and the age estimates}

\begin{figure*}
  \centering
\includegraphics[angle=-90,width=7.5cm, clip=]{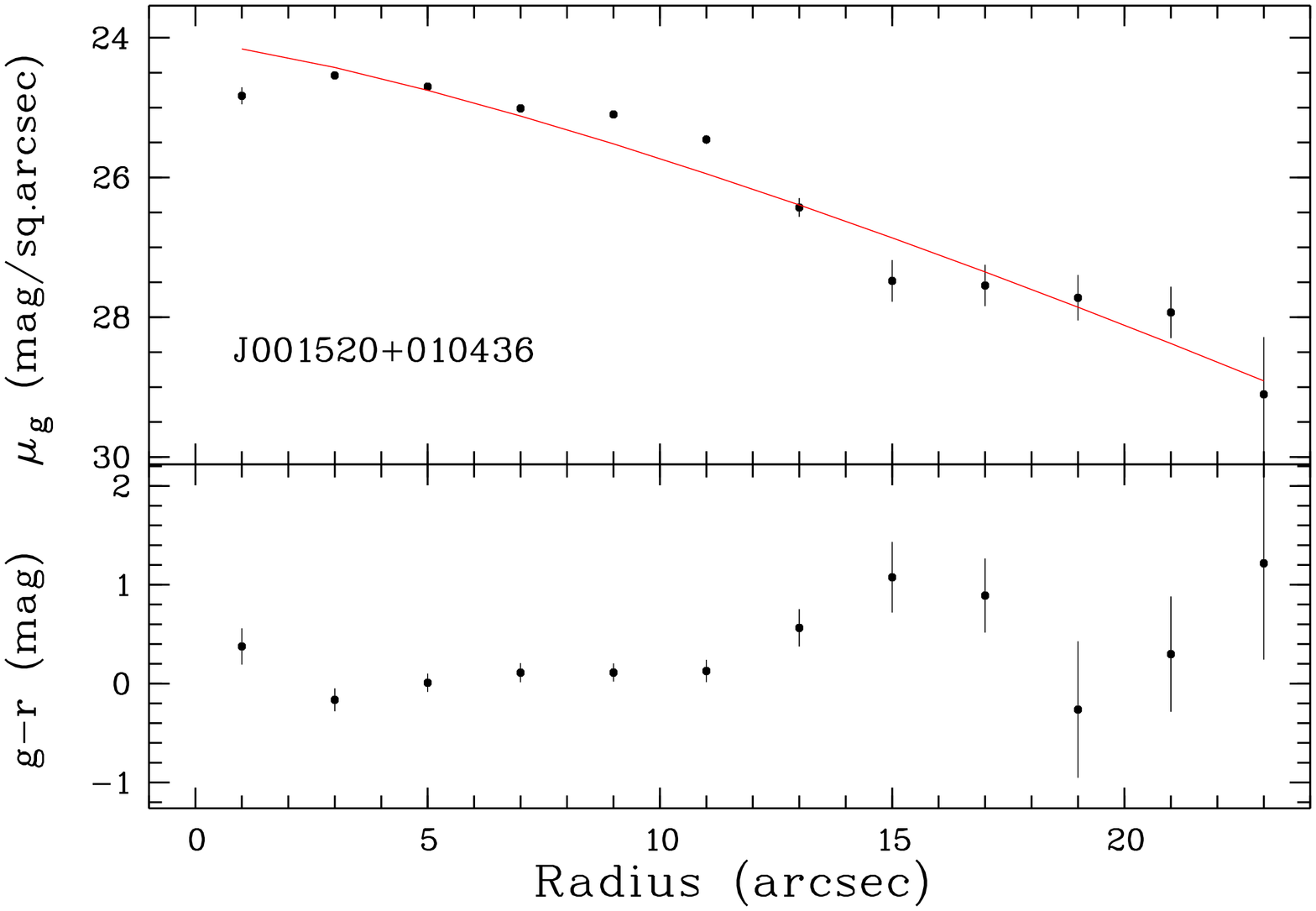}
\includegraphics[angle=-90,width=7.5cm, clip=]{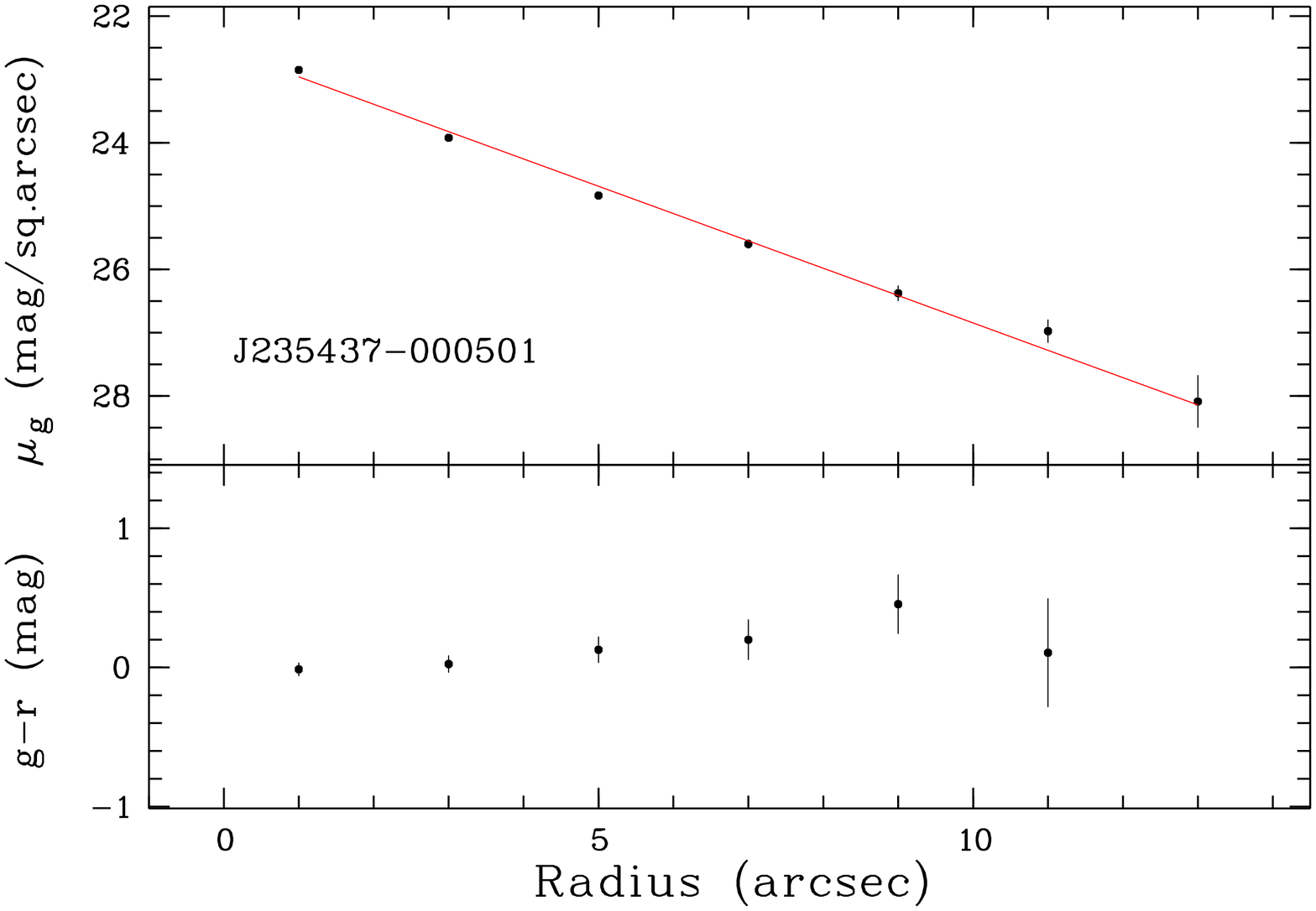}
  \caption{
Left-hand panel: the SB
in $g$ filter and $g-r$ colour of SDSS J0015+0104
versus the effective radius. The solid lines show the S\'ersic model fit
 on the profile part with $<$23 arcsec with exclusion of the very central
depression and two points at R=9 and 11 arcsec as clearly related to
the light of the superimposed \HII\ region.
Right-hand panel: same for galaxy SDSS J2354--0005.
}
	\label{fig:SB_prof}
 \end{figure*}

To construct surface brightness (SB) profiles, we adopted
a geometrical centre
for each galaxy as determined by eye and measured values for concentric round
apertures.
The centre of
J0015+0105 is situated $\sim$10 arcsec  SE from the position of the edge
\HII\ region catalogued as the SDSS emission-line galaxy and whose spectrum
was taken in \citet{Guseva09}. Similarly, for J2354--0005, the centre for
round apertures was put at $\sim$1 arcsec SW from the position of
the emission  knot, catalogued as the SDSS galaxy with the
respective emission-line redshift.
From the SB radial profiles in $g$ filter and the colour
 $(g-r)$  shown  in Figure~\ref{fig:SB_prof}, one can derive the
'mean' colour of the underlying `disc' (excluding outer parts with large
noisy deviations).
For J0015+0104 region within $R = $15 arcsec  this mean $(g-r)$=$\sim$0.15.
Using the transformation formula of \citet{Lupton05}, this translates to
the relation $\mu_{\rm B}$=$\mu_{g}$+0.27~mag~arcsec$^{-2}$.
     Then, from the galaxy SB profile in $g$-filter, the `optical' radius
$R_{\rm opt}$ at $\mu_{\rm B}$=25~mag~arcsec$^{-2}$
and the Holmberg radius R$_{\rm Hol}$ at $\mu_{\rm B}$=26.5~mag~arcsec$^{-2}$
are $R_{\rm opt,eff}$=6.4 arcsec and $R_{\rm Hol,eff}$=12.4 arcsec.
To transform these effective radii to real ones, the correction factor
$(b/a)^{-1/2}$=1.17 should be applied, where the galaxy axial ratio
$b/a$=0.73 was adopted from our photometry. This results in
the `optical' radius of $R_{\rm opt}$=7.5 arcsec ($\sim$1.03~kpc)
and in the `Holmberg' radius of $R_{\rm Hol}$=14.5 arcsec ($\sim$2.0~kpc).
Similar estimates for J2354--00005 with the "mean" $(g-r)$=$\sim$0.10
results in $R_{\rm opt,eff}$=5.0 arcsec and $R_{\rm Hol,eff}$=8.7 arcsec
% R_25=5.0 R_Ho=8.7
% $(b/a)^{-1/2}$=1.17 = sqrt(1/0.563)=1.333 and   scale 0.156 pc in 1.0"
and transformation them to the real values gives $R_{\rm opt}$=6.7 arcsec
($\sim$1.04~kpc) and $R_{\rm Hol}$=11.6 arcsec($\sim$1.8~kpc).

\begin{table}
\caption{Photometric parameters of J0015+0104 and J2354--0005}
\label{tab:photo}
\begin{tabular}{lcc} \\ \hline \hline
Parameter                           & J0015+0105       & J2354--0005      \\ \hline
$g_{\rm tot}$                       & 18.02$\pm$0.01   & 18.52$\pm$0.02   \\
$(u-g)_{\rm tot}$                   & 0.91$\pm$0.03    & 0.58$\pm$0.04    \\
$(g-r)_{\rm tot}$                   & 0.19$\pm$0.01    & 0.10$\pm$0.03    \\
$(r-i)_{\rm tot}$                   & --0.08$\pm$0.01  & 0.13$\pm$0.04    \\
%$(i-z)_{\rm tot}$                   & --0.12$\pm$0.07  & --               \\
$B_{\rm tot}$                       & 18.31$\pm$0.02   & 18.79$\pm$0.03   \\
%$R_{\rm g,25}$ ($\arcsec$)          & 13.8             & 13.8            \\
%$R_{\rm r,25}$ ($\arcsec$)          & 15.7             & 15.7            \\
$(b/a)_{\rm 25}$                     & 0.734            & 0.563           \\
%$R_{\rm g,Hol}$ ($\arcsec$)         & 20.9             & 20.9            \\
%$R_{\rm r,Hol}$ ($\arcsec$)         & 24.3             & 24.3            \\ \hline
%Opt. size (kpc)                    &                  &                  \\
% -----------------------------------------------      -------------
$\mu_{0,g}$(mag~arcsec$^{-2}$)      & 24.04$\pm$0.28   & 22.75$\pm$0.08   \\
$\mu_{0,r}$(mag~arcsec$^{-2}$)      & 24.40$\pm$0.15   & 22.85$\pm$0.10   \\
$\mu_{0,B}$(mag~arcsec$^{-2}$)      & 24.15$\pm$0.32   & 23.01$\pm$0.12   \\
$\mu_{0,B,c,i}$(mag~arcsec$^{-2}$)  & 24.47$\pm$0.32   & 23.55$\pm$0.12   \\
$n_{\rm g}$(S\'ersic)                 & 1.23$\pm$0.24    & 1.00$\pm$0.01    \\
$\alpha_{\rm g}$(arcsec)         & 7.3$\pm$1.8      & 2.7$\pm$0.1      \\
%$n_{\rm r}$(Sersic)                 & 1.13$\pm$0.06    & 1.00$\pm$0.01   \\
%$\alpha_{\rm r}$($\arcsec$)         & 7.1              & 2.7?            \\
$(u-g)_{\rm outer,c}$               & 0.93$\pm$0.08    & 0.79$\pm$0.05    \\
$(g-r)_{\rm outer,c}$               & 0.19$\pm$0.05    & 0.09$\pm$0.03    \\
$(r-i)_{\rm outer,c}$               & 0.15$\pm$0.07    & 0.03$\pm$0.04    \\
$T$(`old' population)               & 4.5$\pm$1.5~Gyr & 2.7$\pm$0.8~Gyr    \\
%$T$(`old' population)              & $\sim$4.5$\pm$1.5~Gyr & $\sim$2.7$\pm$0.8~Gyr    \\
\hline \hline
\multicolumn{3}{p{6.8cm}}{%
% (1)   $(b/a)$-- from NED;
(1) -- All values of the total magnitudes are not corrected
for the foreground MW extinction;
(2) -- colours of the outer parts are corrected for the
foreground MW extinction. }
\end{tabular}
\end{table}

From the above fits (for the internal regions with $R <$12 arcsec  for
J0015+0104  and for 2 $< R < $ 12 arcsec for J2354--0005, excluding
the central SF knot), we also
estimated the central SB. For J0015+0104, the
best fits are for $g$ and $r$ filters, which give $\mu_{\rm g}^0$=24.04 and
$\mu_{\rm r}^0$=24.40~mag~arcsec$^{-2}$ with the scalelength of
7.3 arcsec, that corresponds to the linear scalelength of 1.01 kpc. When
corrected for $(b/a)^{-1/2}$, this gives the
radial disc scalelength of 1.18 kpc.  For J2354--0005, the similar best fits
for $g$ and $r$ filters give $\mu_{\rm g}^0$=22.85 and
$\mu_{\rm r}^0$=22.75~mag~arcsec$^{-2}$, with the scalelength of 2.7 arcsec,
that corresponds to the linear scalelength of 0.42 kpc. When corrected for
$(b/a)^{-1/2}$, this gives the radial disc scalelength of 0.56 kpc.
With the same magnitude transform as above, we derive the {\it observed}
central blue SB of both galaxies: $\mu_{\rm B}^0$=24.14 and
22.87~mag~arcsec$^{-2}$ (after corrections for A$_{\rm B}$=0.11 and 0.14
according to Schlegel et al. 1998).
Due to the galaxy discs inclination, their observed central SB are somewhat
enhanced. We correct the latter values, taking the visible axial ratios
$p=b/a$=0.73 and 0.56, and adopting the internal axial ratio $q=$0.20. The
respective inclination angles $i$, derived from the well known formula:
$\cos(i)^{2} = (p^{2}-q^{2})/(1-q^{2})$,  are $i=$45.1\degr\
and 57.5\degr, respectively.
The respective corrections for $\mu_{\rm B}^0$ are equal to 0.33 and
0.68~mag~arcsec$^{-2}$. Thus, the central SB corrected
for inclination appear as follows: $\mu_{\rm B,0,c,i}$=24.47 and
23.55~mag~arcsec$^{-2}$, and both objects should be classified as
genuine `low surface brightness' galaxies.

The measured $g_{\rm tot}$ magnitudes and respective
% derived, the Galaxy extinction
% $u,g,r,i$ magnitudes and derived, the Galaxy extinction corrected
colours $(u-g), (g-r)$ and $(r-i)$  for the whole galaxies and
the derived $B_{\rm tot}$  as well as the axial ratio $b/a$
are given in the upper part of Table~\ref{tab:photo}.
In the middle part we give the derived central SB in $g$ and $r$ filters,
the respective central brightness in $B$, transformed according to Lupton
at al. (2006), and this parameter corrected for the Milky Way extinction
\citep{Schlegel98} and inclination. Also the model fit parameters
for $g$ filter are included: the Sersic profile index and its characteristic
radius. In the bottom three rows we present the extinction-corrected
$(u-g), (g-r)$ and $(r-i)$ colours for the outer regions of both galaxies.
The last row shows the age estimates of the respective model stellar
populations as derived from the comparison with the model evolutionary
tracks. Both the colours and ages are explained below.

One of our main goals was to compare the observed colours of stellar
population with the {\tiny PEGASE}2 model evolutionary tracks, in order to obtain
the estimates of the maximal stellar ages in these very metal-poor galaxies.
As one can see from the $(g-r)$ colour radial profile of J0015+0104 in
Fig.~\ref{fig:SB_prof} (left panel), due to the contribution
of the mentioned above \HII\ region to the light of the outer parts
of the galaxy, it is difficult to estimate from the profile the outer colours
and to decide
whether there is a colour gradient in the underlying stellar population.

\begin{figure}
  \centering
 \includegraphics[angle=-90,width=9.0cm, clip=]{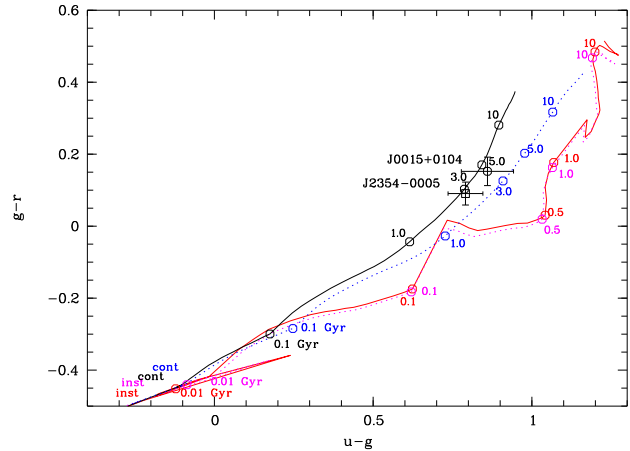}
  \caption{Two squares with error bars show the $ugr$ colours of the outer
parts (see description in the text)
of SDSS J0015+0104 and J2354--0005 in comparison with the PEGASE2
evolutionary tracks (with account for small shift due to galaxies' redshift)
for instantaneous (rightmost) and continuous (leftmost) SF laws. Ticks
along the tracks mark the time (in Gyr) since the beginning of SF episode.
Stellar metallicity is $z$=0.0004, the best model proxy for the real gas
metallicities in these galaxies. The solid lines are for the Salpeter IMF, while
dotted ones -- for Kroupa et al. IMF.
}
	\label{fig:photo}
 \end{figure}

In Fig.~\ref{fig:photo} we compare the derived colours of the outer
parts for J0015+0104 and J2354--0005 (squares with error bars) from
Table~\ref{tab:photo} with the model tracks from the PEGASE2 package \citep{pegase2}.
To derive the upper limit of the visible stellar population ages,
we exploit the evolutionary tracks for the two extreme SF laws: instantaneous
SF burst and continuous SF with constant SFR. For both SF laws, we show
tracks with the Salpeter IMF (solid) and with the IMF of Kroupa et al.
(dotted line) for metallicity z=0.0004, which is the best proxy
for the O/H values of the galaxies in question. The integrated colours of
J0015+0104 fall in between the Salpeter and Kroupa et al. IMF continuous
SF law tracks. Its rather blue colours $(u-g),(g-r)$
correspond to ages of $\sim$3 Gyr. The square with error bars, corresponding
to colours of the outer parts (obtained by integrating the light in 14 small
round apertures at the mean radial distance of 11 arcsec or
$\sim$1.5~kpc) falls
also in between the Salpeter and Kroupa et al. continuous  SF law tracks,
but in the region with ages of
$\sim$4.5$\pm$1.5~Gyr, in difference with the ages of $\sim$10--13 Gyr,
corresponding to red colours of the  majority of late-type dwarf
galaxies, for which the similar data are acquired (e.g.
\citet{Parodi02}, \citet{Bergvall10}, \citet{Kniazev09}).
For J2354--0005, the total colours appear somewhat bluer
than the model tracks. This can be related to the significant contribution
of the central starburst
emission. The colours of the outer
parts, obtained by integrating the light in five small round apertures
at the mean radial distance of 4.4 arcsec  ($\sim$0.7~kpc)
are redder in $(u-g)$. The position of the respective square is closer
to the Salpeter IMF continuous SF track, with the mean age of
$\sim$2.7$\pm$0.8~Gyr.
In the $gri$ colour plot (not shown) both
tracks go too close each to other, so the galaxy $gri$ colours alone would
be inconclusive. However, the measured $(r-i)$ colours are consistent with
the above conclusions derived from $ugr$ colours.

\section[]{DISCUSSION}
\label{sec:dis}

In Table~\ref{tab:param} we present the main parameters of the studied
galaxies. Since the LSBD galaxy J0015+0104  is not properly identified
in the SDSS, we adopt for its  coordinates those of its geometrical
centre, instead of the \HII\ region located at the edge of the galaxy.
For J2354--0005, however, we keep the SDSS coordinates of the emission-line
knot, which differ from the main body centre only by $\sim$1 arcsec.
From the total magnitudes in filters $g$ and $r$ (Table
\ref{tab:photo}), with the transformation equations of \citet{Lupton05},
we derived the total $B$-band magnitudes, $B_{\rm tot}$=18.31$\pm$0.03 and
18.79$\pm$0.03, respectively. For their adopted distance moduli
$\mu$=32.27 (D=28.4 Mpc) and $\mu$=32.53 (D=32.1 Mpc),
their absolute magnitudes are M$_{\rm B}^0 = -$14.07 and --13.88. The latter
correspond to $L_{\rm B}$=6.55 and 5.47$\times 10^{7}~L_{\rm B}$\sunn.
Then, from the $M$(\HI) derived in the previous section, one obtains the
ratios $M$(\HI)/$L_{\rm B} \sim$2.35 and $\sim$2.2 (in solar units).

From the \HI\ linewidths at 20\%-level of the maximal intensity,
$W_\mathrm{20}$, one can estimate the maximal rotational velocities, using
the standard approximation, as, e.g., formula 12 from \citet{Tully85}
or in \citet{Tully08}.
If one assumes that the planes of \HI\ and stellar discs are close,
the inclination corrections are $1/\sin(i) =$1.4 and 1.18. For the measured
linewidths $W_\mathrm{20}$=29.4~\kms\ and 68~\kms, with the account for
the velocity dispersion of \HI\ gas,  $\sigma_{\rm V}=$8~\kms,
the derived  $V_{\rm rot}\sim$5~\kms\ [\citet{Tully85}, since for
\citet{Tully08} the value $W_\mathrm{20} <$38~\kms\ implies zero rotation
velocity] and 24.3~\kms.
Respectively, the inclination corrected velocities are:
$V_{\rm rot}=$7~\kms\
and 28.6~\kms. The former value is substantially smaller than $V_{\rm rot}$
typical of dwarf galaxies with comparable M$_{\rm B}$ in the faint dwarf
sample Faint Irregular Galaxies GMRT Survey (FIGGS) of \citet{Begum08b}. 
Indeed, the recent \HI\ mapping with GMRT
(Chengalur et al., in preparation) reveals that the very small
width
of \HI\ profile in galaxy J0015+0104 is due to almost face-on inclination
of its \HI\ disc. Therefore, the correct deprojection should increase the
apparent rotation velocity amplitude by several times.
Having $V_{\rm rot}$ and the characteristic size of the galaxy, one can
estimate its total (dynamical) mass which is necessary to balance the
centrifugal force within a certain radius. Due to the mentioned above
situation (which is quite rare), for J0015+0104 such estimate, based only
on the NRT profile and the optical body inclination, is highly unreliable.
We refer to the GMRT data analysis in preparation for its better dynamical
mass estimate.

The typical radii of \HI\ discs (at the column  density level of
10$^{19}$~atoms~cm$^{-2}$) in dwarf galaxies with $M_{\rm B} \sim$--13,
close to those of the studied LSBDs, are 2.5--3 times larger than the
Holmberg
radius \citep[e.g.,][]{Begum08a,Begum08b}. Therefore, we accept that the
\HI\ radius of J2354--0005 is $R_{\rm HI} =$ 5.3~kpc.
%%  R_Holm(eff) = 8.7", after correct. for sqrt(a/b) => 11.5"
Then, from the relation
$M(R < R_{\rm HI}$) = $V_{\rm rot}^2~\times$~$R_{\rm HI}/G$, where $G$ is
the gravitational constant, one derives the total mass within $R_{\rm HI}$
as follows:
$M_{\rm tot}$=10.1$\times$10$^{8}$~M\sunn.
%  M=3.0^2*10^12*5.3*3.1*10^21/(6.67*10^(-8)) = 22.2*10^{41} g = 11.1*10^8 Mo
%  M=2.86^2*10^12*5.3*3.1*10^21/(6.67*10^(-8)) = 20.15*10^{41} g = 10.1*10^8 Mo
To get the estimates of the gas mass in both galaxies, we sum $M$(\HI)
and $M$(He) (a fraction of 0.33 of \HI\ mass) that gives, respectively:
$M_{\rm gas}$=2.05$\times$10$^{8}$~M\sunn\ and 1.85$\times$10$^{8}$~M\sunn.

The mass of stars can be estimated as follows. We take from
Table~\ref{tab:photo} the total $g$ magnitude of J0015+0104 equal to 18.02.
This corresponds to the extinction-corrected absolute magnitude
M$_{\rm g}$=--14.35. Following the prescripts of
\citet{Zibetti09}\footnote{We choose this prescripts as the most advanced
and, in contrast with previous works, incorporating the SF history at
relatively late epochs, which better corresponds to LSBD galaxies
considered here.}, we use
the mass-to-light ratio $\Upsilon$ based on $g$-filter luminosity and the
extinction-corrected colour $(g-i)_{\rm tot}$=0.06$\pm$0.04. This reads as
$\log~\Upsilon_{\rm g} = -1.197+1.431\times(g-i)$.
Hence  $\log~\Upsilon_{\rm g}$= --1.111$\pm$0.057,
and  $\Upsilon_{\rm g}$ is  0.077$\pm ^{0.011}_{0.009}$.
For the absolute magnitude of J0015+0104 of $M_{\rm g}=-$14.35 and the
absolute $M_{\rm g, \odot}=$5.15, the galaxy luminosity appears
$L_{\rm g,J0015}$ = 6.3$\times$10$^{7}$~$L$\sunn.
% M_sun(g)=5.15    M_g=-14.35  delta_M_g=19.50, that is 6.3*10^7 L_sun
% that corresponds to Mass =0.945*10^7 Mo
With $\Upsilon_{\rm g}$=0.077 from above, this implies the total
stellar mass of
(0.048$\pm ^{0.007}_{0.006}$)$\times$10$^{8}$~M\sunn.
Then the baryonic mass is
$M_{\rm bary}$=(2.10$\pm$0.13)$\times$10$^{8}$~M\sunn\ and the
gas mass-fraction
$f_{\rm gas}$=$M_{\rm gas}$/($M_{\rm gas}$+$M_{\rm star}$)=0.976.
The similar estimate of the stellar mass for galaxy J2354--0005,
based on its
% not ext.correcct. g-i=0.23+-0.05  E(g-i)=>0.06  g-i_0=0.17
% gives log(Upsilon)=-0.968+-0.072  [ -0.896 and -1.040]
% -14.04=-32.53+18.49
$(g-i)_{\rm tot}$=0.17$\pm$0.05, uses value
of $\Upsilon_{\rm g}$=0.108$\pm^{0.019}_{0.017}$.
With its absolute magnitude of $M_{\rm g}=-$14.05, its luminosity is
$L_{\rm g,J2354}$ = 4.7$\times$10$^{7}$~$L$\sunn.
Then its total stellar mass is (0.051$\pm$0.008)$\times$10$^{8}$~M\sunn.
% M_sun(g)=5.15    M_g=-14.05  delta_M_g=19.28, that is 5.152*10^7 L_sun
The respective baryonic mass is
$M_{\rm bary}$=1.90$\times$10$^{8}$~M\sunn, and $f_{\rm g}$=0.973.

The ratio of $M_{\rm tot}$ [inside $R$(\HI)]  and $M_{\rm bary}$
% $M_{\rm tot}$=11.1$\times$10$^{8}$~M\sunn\ and the total baryon
% mass, M$_{\rm bary}$=1.90$\times$10$^{8}$~M\sunn.
for J2354--0005 is $\sim$5.3, rather close to the cosmic mean value derived
for the best fit $\Lambda$CDM cosmology.
For galaxy J0015+0104, if we adopt its rotation velocity as derived with
the inclination correction for its optical body, and similar \HI\ radius,
such ratio will be $\sim$0.2,
while by the definition this should not be less than 1.
The latter directly indicates that the \HI\ disc is much closer to face-on.
Even in the case when the ratio $R$(\HI)/$R_{\rm Holm}$ is significantly
larger than the adopted here (2.75), that is similar to those of a few very
gas-rich galaxies, with the ratio $R$(\HI)/$R_{\rm Holm}$  of 5--10 [e.g.,
as in And~IV and NGC~3741, \citet{Begum08a} and KK~246, \citet{Kreckel11a}],
the estimate of $M_{\rm tot}$ within $R$(\HI) will be still smaller than
$M_{\rm bar}$.  Anyway, to get a better quality estimate of the DM halo mass
and its radial distribution, one needs in \HI\ mapping of the galaxy.
In Table \ref{tab:param}, along with the summary of the main parameters for
SDSS J0015+0104 and J2354--0005, we remind the properties of SDSS
J0926+3343 with 12+$\log$(O/H)=7.12 \citep{J0926}, in order to emphasize
the range of other parameter diversity in two the most metal-poor LSB dwarf
galaxies known. As one can see, the difference of their main parameters
(gas mass, corrected central SB, blue luminosity), does not exceed a factor
of 2--3. Their O/H and very high gas mass-fractions are very close;
the ages of the main visible old population being non-cosmological,
can differ in several times, or can be about the same if $T \sim$3~Gyr.

While we consider these two LSBDs and several of their counterparts in
the Lynx-Cancer void as unusual, it is reasonable to
illustrate properties of more or less typical LSBDs. One of the best
LSBD samples is a subsample of late-type galaxies in the Local Volume and its
environment studied by \citet{vZee06}. We made simple statistical analysis
of that subsample properties (which includes UGC2684, UGCA20, UGC3174,
UGCA357, UGC891, UGC5716, UGC300, UGC11820 and UGC634) and summarize their
medians in the last column of Table \ref{tab:param}. They appear quite similar
on luminosity, the central SB (as expected from their definition), and
$M$(\HI)/$L_{\rm B}$ ratio. However, their O/H typically is four to eight times higher
and their gas mass-fraction corresponds to significantly larger proportion
of stellar mass. The latter is related to their redder integrated colours.
Thus, one can conclude that the main differences between the LSBDs studied
in this paper and more typical ones are their too low metallicity (in
comparison to the expected for their luminosity, see Papers II and III) and
too blue integrated and the outer region colours.

Such unusual LSBDs which combine the very low metallicity, very
high gas mass fraction and blue colours of stellar population in galaxy outer
parts are extremely rare. So far they are found mostly in voids \citep[and probably
sometimes at the periphery of groups, Paper III and]
[]{CP2013}.
A few galaxies with $f_{\rm gas} \sim$0.9 found in \citet{Geha06} and
\citet{McGaugh2010} can appear as similar objects, but this needs a careful
study.
Three very gas-rich dwarf galaxies also were
recently found near void centres in the Void Galaxy Survey \citep{Kreckel12}.
Therefore it is natural to look for surroundings of galaxies in the
question.
The examination of the immediate environment of SDSS J0015+0104
with NED shows no catalogued  galaxy with known radial velocity
within the projected distance of $\sim$2~Mpc and $\Delta V \sim$500~\kms.
The careful study of the neighbouring sky region reveals no {\it luminous}
galaxies (with $M_{\rm B} < $--19) at the projected distances closer
than 7~Mpc with $\Delta V <$ 500~\kms. This is consistent with the separation
of this cell of the local Universe as a part of the "Eridanus void"\,
according to the description by \citet{Fairall98}.
In his Table~4.1, the Eridanus void is centred on the sky at R.A.$\sim$1.0~h,
Dec.$\sim$0\degr\ and at the distance, corresponding to
$V_{\rm LG} \sim$2500~\kms.
The similar inspection of J2354--0005 surroundings with NED shows that it
also resides within the void but closer to its border. The nearest luminous
galaxy NGC~7716 is at $\sim$2.5~Mpc in projection, with $\Delta V$=260~\kms,
that is at the physical distance of 4.3~Mpc.
The nearest subluminous galaxy UGC~12857 is located at the projected
distance of 0.86~Mpc and $\Delta V$=166~\kms, or at the physical distance
of 2.4~Mpc.
% The claimed size of this void corresponds to $\sim$4000~\kms, or $\sim$55~Mpc.
%The void description should be improved, incorporating the new galaxy velocity
%data and the advanced methods of void separation. But within this earlier
%scheme, J0015+0104 is situated within $\sim$4 Mpc from the centre of the
%void with R$\sim$27~Mpc.

The two very metal poor LSBDs discussed above, are similar to several
of the most metal poor LSBDs in the Lynx-Cancer void, described
in Paper III. The very high concentration of such `unevolved' dwarfs in
that void allowed us to conclude that the very
low-density environment provides the special conditions for (a part of)
low-mass galaxy formation and evolution. The existence in the Eridanus void
of these two unusual extremely metal defcient (XMD) LSBDs,
as well as of several other XMD galaxies (work in progress)
is consistent with a more general statement on the causal relations between
voids and `unevolved' dwarf galaxies.

% SDSS J001521.27+010428.4

\begin{table*}
\caption{Main parameters of SDSS J0015+0104, J2354--0005, J0926+3343 and median for 9 typical LSBD sample}
\label{tab:param}
\begin{tabular}{lcccc} \\ \hline
Parameter                          & J0015+0104             &  J2354--0005           &  J0926+3343             &  `typical' LSBD$^{(8)}$      \\ \hline
R.A.(J2000.0)                      & 00 15 21.22            & 23 54 37.30            & 09 26 09.45             &      --              \\
DEC.(J2000.0)                      & $+$01 04 29.9          & $-$00 05 06.1          & $+$33 43 04.1           &      --              \\
$A_{\rm B}$ (from NED)             & 0.11                   & 0.14                   & 0.08                    &      --              \\
$B_{\rm tot}$                      & 18.31$^{(1)}$          & 18.79$^{(1)}$          & 17.34$^{(6)}$           &      --              \\
$V_{\rm hel}$(HI)(\kms)            &2035$\pm$1$^{(1)}$      & 2310$\pm$4$^{(1)}$     & 536$^{(6)}$             &      --              \\
$V_{\rm LG}$(HI)(\kms)             &2207                    & 2489                   & 488$^{(6)}$             &      --              \\
Distance (Mpc)                     &28.4$^{(1)}$            & 32.1$^{(1)}$           & 10.7$^{(6)}$            &      16.0            \\
$M_{\rm B}^0$                      & --14.07$^{(1)}$        & --13.88$^{(1)}$        & --12.90$^{(6)}$         &    --15.53           \\
Opt. size (arcsec)$^{(3)}$        & 7.5$\times$5.5$^{(1)}$ & 6.5$\times$3.6$^{(1)}$ & 35.8$\times$9.9$^{(6)}$ &      --              \\
Opt. size (kpc)                    & 1.03$\times$0.75       & 1.01$\times$0.56       & 0.93$\times$0.26$^{(6)}$&      --              \\
$\mu_{\rm B,c,i}^0$(mag~arcsec$^{-2}$) & 24.47  $^{(1)}$    & 23.55$^{(1)}$          & 25.4$^{(6)}$            &     23.70            \\
12+$\log$(O/H)                     & 7.07$\pm$0.07$^{(7)}$  & 7.36$\pm$0.13$^{(7)}$  & 7.12$^{(6)}$            &      8.0             \\
\HI\ int.flux$^{(4)}$              & 0.81$\pm$0.04$^{(2)}$  & 0.50$\pm$0.04$^{(2)}$  & 2.54$^{(6)}$            &      --              \\
$W_\mathrm{50}$ (km s$^{-1}$)      & 22$\pm$2$^{(2)}$       & 38$\pm$7$^{(2)}$      & 47.4$^{(6)}$            &      --              \\
$W_\mathrm{20}$ (km s$^{-1}$)      & 32$\pm$3$^{(2)}$       & 70$\pm$12$^{(2)}$      & 80.5$^{(6)}$            &      --              \\
$V_\mathrm{rot}$ (HI)(\kms)        & 7::$^{(2)}$            & 29$^{(2)}$             & 32$^{(6)}$              &      --              \\
$M$(\HI) (10$^{7} M$\sunn)         & 15.4$^{(1)}$           & 12.2$^{(2)}$           & 6.8$^{(6)}$             &     45.6             \\
$M_{\rm dyn}$ (10$^{7} M$\sunn)    & --                     & 101$^{(1)}$            & 124$^{(6)}$             &     339              \\
$M$(\HI)/$L_{\rm B}^{(5)}$         & 2.35$^{(1)}$           & 2.2$^{(1)}$            & 3.0$^{(6)}$             &     2.2              \\
$f_{\rm gas}$                      & 0.98$^{(1)}$           & 0.97$^{(1)}$           & 0.98$^{(1)}$            &     0.89             \\
$T$(main star population)          & 3--6~Gyr$^{(2)}$       & 0.9--2.5~Gyr$^{(2)}$   & 1--3~Gyr$^{(6)}$        &     10~Gyr           \\
\hline
\multicolumn{5}{p{13.2cm}}{(1) -- derived in this paper; (2) -- derived
from NRT \HI\ profile; (3) -- $a \times b$ at
$\mu_{\rm B}=$25\fm0~arcsec$^{-2}$; (4) -- in units of Jy$\cdot$\kms; (5) --
in solar units; (6) \citet{J0926}; (7) \citet{Guseva09};
(8) median values for LSBG sample from \citet{vZee06}.
}
\end{tabular}
\end{table*}

Summarising the results and discussion above, we draw the following
conclusions:

\begin{enumerate}
\item
The two genuine LSBD galaxies SDSS J0015+0104 and J2354--0005 with
the extremely low metallicities are situated deeply in the Eridanus void,
with distances to the nearest luminous galaxies of $D_{NN}$ of $>$7 and
4.3 Mpc, respectively.
\item
The \HI\ integrated flux of these galaxies, along with the SDSS-based optical
photometry, indicate that they are very gas-rich objects, with
$M$(\HI)/L$_{\rm B} \sim$2.35 and $\sim$2.3, and with the derived gas
mass-fractions of $f_{\rm gas} \sim$0.98 and 0.97, respectively.
\item
For J2354--0005, the total mass estimate based on the galaxy $W_{\mathrm 20}$,
size, and inclination angle derived from the optical axial ratio, leads
to $M_{tot}/M_{bary} \sim$5.3.
For J0015+0104, having very narrow \HI\ profile, the related low rotation
velocity estimate is due to the large misalignment between the optical and
\HI\ discs. The orientation of the latter (as evident from the GMRT
\HI\ mapping) is close to face-on.
\item
The $(u-g)$, $(g-r)$ and $(r-i)$ colours of SDSS J0015+0104 and J2354--0005
are rather `blue'. The `outer' region colours show no tracers of ubiquitous
old stellar population with ages of 10--12~Gyr. They well match the PEGASE2
model track for the evolving stellar population with continuous SF for ages
of T$\sim$4.5 Gyr for the former and T$\sim$2.7~Gyr for the latter. Thus,
all three observational parameters: O/H, $f_{\rm g}$ and blue colours
consistently evidence for evolutionary young status of these dwarfs.
\item
Void-type environment seems to favour the appearance of unusual dwarf
galaxies with properties of unevolved matter (very low metallicity and very
high gas mass fraction). Therefore, the dedicated search for such objects
among void sample galaxies can be one of the most efficient means.

\end{enumerate}

\section*{Acknowledgements}

SAP and YAL acknowledge the partial support of this project through the
RFBR grant No.~11-02-00261 and the Russian Federal Innovation Program
(contract No.~14.740.11.0901 and proposal No.~2012-1.5-12-000-1011-004).
SAP and JMM acknowledge the NRT TAC for
allocation of TOO for this program in 2011-2012. This work was started
during SAP visit to Observatoire de Paris, GEPI in 2011. He thanks it
for support and hospitality. AYK acknowledges the support from the
National Research Foundation (NRF) of South Africa. The questions and
suggestions of the anonymous reviewer helped to improve the paper quality.
The authors acknowledge the spectral and photometric data and the related
information available in the SDSS database used for this study.
The Sloan Digital Sky Survey (SDSS) is a joint project of the University of
Chicago, Fermilab, the Institute for Advanced Study, the Japan Participation
Group, the Johns Hopkins University, the Max-Planck-Institute for Astronomy
(MPIA), the Max-Planck-Institute for Astrophysics (MPA), New Mexico State
University, Princeton University, the United States Naval Observatory, and
the University of Washington. Apache Point Observatory, site of the SDSS
telescopes, is operated by the Astrophysical Research Consortium (ARC).
This research has made use of the NASA/IPAC Extragalactic
Database (NED), which is operated by the Jet Propulsion Laboratory,
California Institute of Technology, under contract with the National
Aeronautics and Space Administration.

%===========================================================================

\bsp

\label{lastpage}

\end{document}